\documentclass[12pt,english]{paper}
\usepackage[T1]{fontenc}
\usepackage[latin9]{inputenc}
\usepackage[a4paper]{geometry}
\usepackage{color}
\usepackage{babel}
\usepackage{amsmath}
\usepackage{graphicx}
\usepackage{esint}
\usepackage[numbers]{natbib}
\usepackage[unicode=true,pdfusetitle,
 bookmarks=true,bookmarksnumbered=false,bookmarksopen=false,
 breaklinks=false,pdfborder={0 0 1},backref=false,colorlinks=true]
 {hyperref}
\hypersetup{
 citecolor=blue, linkcolor=black, urlcolor=blue}

\makeatletter
\numberwithin{equation}{section}

\makeatother

\begin{document}

\title{Intrinsic Thickness of QCD Flux-Tubes }

\author{Vikram Vyas\\
}

\institution{Physics Department, St. Stephen's College\\
Delhi University, Delhi 110007, India\\
Email:visquare@gmail.com}
\maketitle
\begin{abstract}
The notion an intrinsic thickness of a QCD flux-tube is explored heuristically
with in the framework of Gauge/Gravity Duality. 
\end{abstract}

\section{Introduction}

In the past few years the lattice simulation of QCD flux-tubes has
reached a level of precision where they can be unambiguously compared
with the expectations from various effective string descriptions (two
of the most recent investigations are \citep{HariDass:2006pq,Athenodorou:2009fk},
and a lucid review with a history of numerical simulations of flux-tube
using lattice gauge theory is given in \citep{Teper:2009uq}.) These
results show that the ground state energy of a long flux-tube matches
very well with that obtained by Arvis \citep{Arvis:1983fp} for the
Nambu-Goto string in four flat dimensions. It was also pointed out
in \citep{Athenodorou:2009fk}, that there are few excited states
of a closed flux-tube that do show a large deviation from the Nambu-Goto
prediction. Excited states of open flux-tube have only been calculated
in three dimensions \citep{Brandt:2009tc} and do not show any unambiguous
deviations from the predictions of the Nambu-Goto string, within the
accuracy of the simulations.

The effective string theory description of QCD flux-tubes assumes
that the length of the flux-tube, $L,$ is much greater than any intrinsic
thickness, $l_{w}$, of the flux-tube. Space-time history of such
a flux-tube, in the absence of dynamical quarks, can be approximated
by a two-dimensional surface or the world-sheet. The dynamics of this
world-sheet is independent of the two-dimensional coordinate system
used to describe it. A general reparameterization invariant action
can be organized into relevant, marginal and irrelevant terms by dimensional
analysis. The relevant term, which is proportional to the area of
the world-sheet, is the Nambu-Goto action. Then there are marginal
and irrelevant terms constructed from various powers of intrinsic
and extrinsic curvatures (see for e.g. \citep{David:1990ju}). 

The notion of an intrinsic thickness of a flux-tube, though well motivated,
is imprecise in QCD as we do not understand their dynamical origin.
It is here that one expects that the conjectured gauge/gravity duality
\citep{Polyakov:1998aa,Polyakov:1998ju,Aharony:2000aa,Horowitz:2006ct}
should allow us to delineate the effects of the intrinsic thickness
of the flux-tube. A fundamental feature of the gauge/gravity duality
is that the strings dual to the gauge theory live in (at least) five-dimensional
curved space. These five-dimensional strings then should provide an
\emph{exact} description of the four-dimensional QCD flux-tubes. These
fundamental strings have at least one additional degree of freedom
as compared to the effective four-dimensional strings. It is tempting
to associate this additional degree of freedom with the intrinsic
thickness of a QCD flux-tube. This can be made more precise in the
context of AdS/CFT correspondence \citep{Aharony:2000aa,Susskind:1998dq,Polchinski:2001ju},
by looking at the size of the holographic projection of a five-dimensional
object on to the four-dimensions. 

AdS/CFT correspondence \citep{Maldacena:1997re} can be regarded as
an example of a more general gauge/gravity duality \citep{Horowitz:2006ct}
according to which all gauge theories, including QCD, have a dual
description in terms of fundamental strings living in a higher-dimensional
curved space-time. For QCD we do not known the precise geometry in
which these putative five-dimensional strings live, but an approximate
description which captures some of the properties of the QCD can be
obtained by modifications of $AdS_{5}$ space. A particularly simple
modification was used in \citep{Polchinski:2001ju} to understand
the form factors of hadrons in terms of the intrinsically thick effective
QCD strings arising from the holographic projection of the fundamental
five-dimensional strings. We will use their modification to give a
meaning to the idea of an intrinsic thickness of a flux-tube.

\section{Flux-tube from a Fundamental String\label{sec:Flux-tube-from-a}}

One way of stating the conjectured gauge/gravity duality is via the
expectation value of a Wilson loop. The expectation value of a Wilson
loop in a four-dimensional Yang-Mills gauge theory is formally given
by
\begin{eqnarray}
W[\Gamma,A] & = & \mathrm{Tr}\hat{P}\exp\left\{ i\oint_{\Gamma}A\right\} ,\nonumber \\
<W[\Gamma]>_{YM} & = & \int[DA]\exp\left\{ -S_{YM}[A]\right\} W[\Gamma,A],\label{eq:WilsonLoop}
\end{eqnarray}
where $\Gamma$ is a non-intersecting closed loop in four-dimensions,
$A$ represents the matrix valued vector potential of the $SU(N)$
Yang-Mills theory, which have been path-ordered along the curve $\Gamma$
using the path ordering operator $\hat{P}$, and $S_{YM}[A]$ is the
Euclidean Yang-Mills action for the gauge fields. The gauge/gravity
duality then implies that this expectation value is also given by
a weighted sum over all the surfaces whose boundary is the curve $\Gamma$
which lives in the flat four-dimensional space, while the surfaces
themselves live in a curved five-dimensional space. Symbolically
\begin{equation}
<W[\Gamma]>_{YM}=\int[DX]\exp\left\{ -T_{o}\int d^{2}\sigma\sqrt{\gamma[X]}\right\} ,\label{eq:gauge/Gravity}
\end{equation}
where
\begin{equation}
X^{m}=X^{m}(\sigma^{0},\sigma^{1}),\; m=1,\ldots,5\label{eq:stringSurface}
\end{equation}
 is the string surface parametrized by the parameters $\left\{ \sigma^{0},\sigma^{1}\right\} $,
$\gamma$ is the determinant of the induced metric
\begin{equation}
\gamma_{ab}=g_{mn}\frac{\partial X^{m}}{\partial\sigma^{a}}\frac{\partial X^{n}}{\partial\sigma^{b}},\label{eq:inducedMetric}
\end{equation}
and $T_{0}$ is the string-tension. The curved five-dimensional space
is described by the metric $g_{mn}$, which following \citep{Polchinski:2001ju}
will be taken to be of the following form

\begin{multline}
ds^{2}=g_{mn}dx^{m}dx^{n}=F(x_{5})\left(dx_{4}^{2}+dx_{1}^{2}+dx_{2}^{2}+dx_{3}^{2}+dx_{5}^{2}\right)\\
=F(y)(dt^{2}+dx_{1}^{2}+dx_{2}^{2}+dx_{3}^{2}+dy^{2}),\label{eq:confiningMetric-1}
\end{multline}
where the function $F(y)$ is assumed to have a minimum at $y=y^{*}$
and for $y\rightarrow0$ it approaches the $AdS_{5}$ limit $ $
\begin{equation}
F(y)\approx\frac{R^{2}}{y^{2}}.\label{eq:boundaryMetric}
\end{equation}

Consider now a Wilson loop made of the world-lines of a quark at the
origin and an anti-quark placed along the $x_{1}$ axis at a coordinate
distance of $L.$ Working in the static gauge
\begin{equation}
\sigma^{0}=\tau=X_{4}=t;\;\sigma^{1}=\sigma=X_{1}=x,\label{eq:staticGauge}
\end{equation}
the classical world-sheet, or the minimal surface, is of the form:
\begin{equation}
X_{c}=(\tau,\sigma,0,0,Y_{c}(\sigma)).\label{eq:minSurface}
\end{equation}
$Y_{c}(\sigma)$, which is a geodesic in the five-dimensional curved
space, satisfies the equation
\begin{equation}
\frac{\partial}{\partial\sigma}\left(\frac{F[Y]Y^{'}}{\sqrt{(1+Y^{'2})}}\right)=F^{'}[Y]\sqrt{(1+Y^{'})^{2}}.\label{eq:EOM}
\end{equation}

\begin{figure}
\includegraphics[scale=0.5]{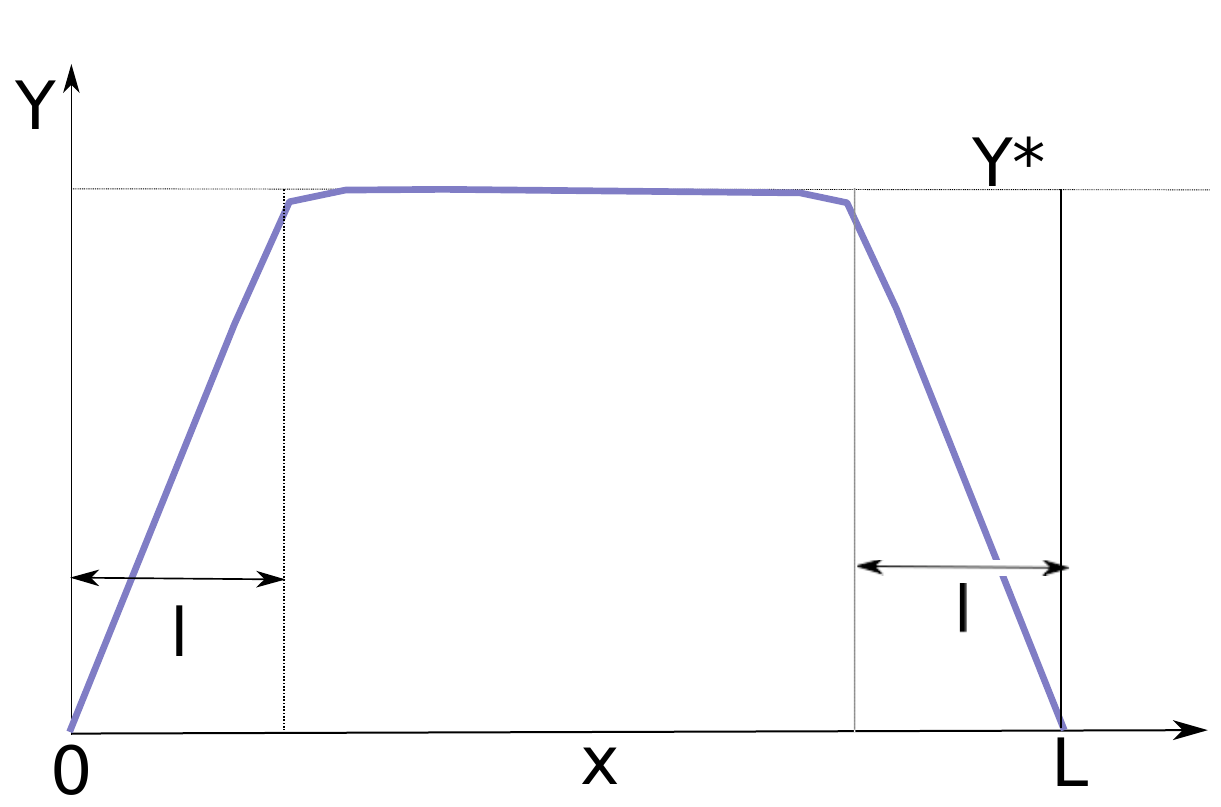}

\caption{$Y_{c}(x)$\label{fig:1}}

\end{figure}

With the assumed form of $F[y]$, qualitatively $Y_{c}(\sigma)$ looks
like the curve shown in Fig.(\ref{fig:1}). In particular one can
approximate
\begin{equation}
Y_{c}(x)=Y^{*}\quad(l\le x\le L-l).\label{eq:hardWall}
\end{equation}
The parameter $Y^{*}$ and $l$ will depend on the precise form of
the confining metric. The corresponding classical action is then given
by
\begin{eqnarray}
S[Y_{c}] & = & T_{0}\int_{-\infty}^{\infty}dt\int_{0}^{l}dxF[Y_{c}](1+Y_{c}^{'2})^{\frac{1}{2}}\nonumber \\
 & + & T_{0}F[Y^{*}]\int_{-\infty}^{\infty}dt\int_{l}^{L-l}dx\nonumber \\
 & + & T_{0}\int_{-\infty}^{\infty}dt\int_{L-l}^{L}dxF[Y_{c}](1+Y_{c}^{'2})^{\frac{1}{2,}}.\label{eq:classicalAction}
\end{eqnarray}
Let us define a position dependent tension,
\begin{equation}
T(x)=\begin{cases}
T_{0}F[Y_{c}](1+Y^{'2})^{1/2} & 0\le x\le l\\
T_{0}F[Y^{*}] & l<x<L-l\:,\\
T_{0}F[Y_{c}](1+Y^{'2})^{1/2} & L-l\le x\le L
\end{cases}\label{eq:variableStringTension}
\end{equation}
and write the classical action as
\begin{equation}
S_{c}=\int_{-\infty}^{\infty}dt\int_{0}^{L}dxT(x).\label{eq:classicalFluxTube}
\end{equation}

\begin{figure}
\includegraphics[scale=0.75]{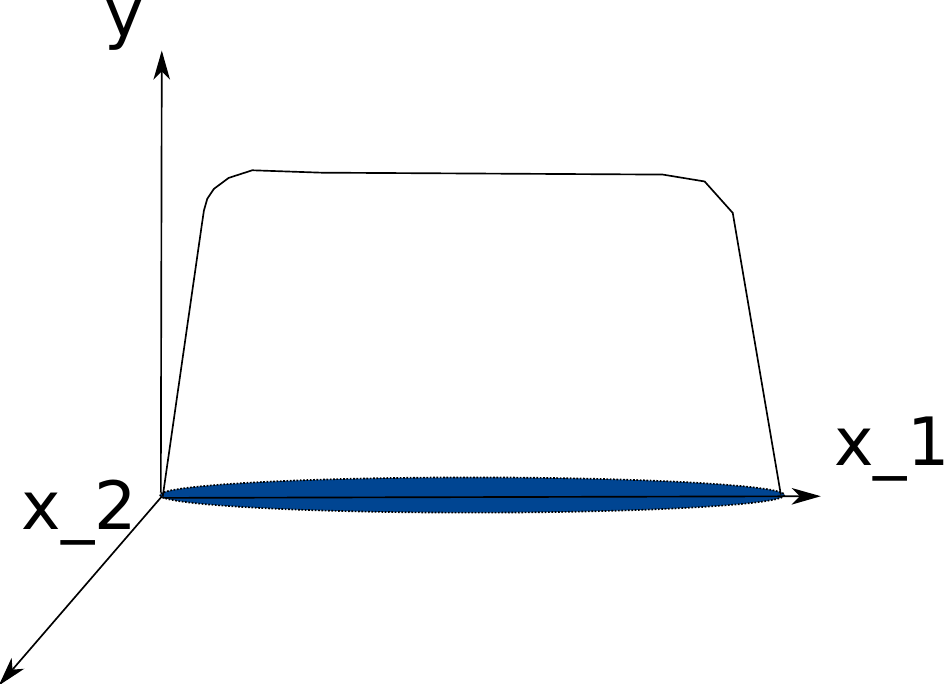}

\caption{Flux-tube from a string\label{fig:Flux-tube}}

\end{figure}

The above action can be thought of as an action for a string along
the $x$ axis in four-dimensions, but with a position dependent string
tension. If one considers a flux-tube like the Nielsen-Olesen vortex
line \citep{Nielsen:1973cs} then one finds that the string tension
is related to the size of the flux-tube by 
\begin{equation}
T\sim\frac{1}{\lambda^{2}},\label{eq:fluxTubeSize}
\end{equation}
 where $\lambda$ characterizes the region beyond which the magnetic
field is essentially zero. Therefore a suggestive interpretation of
\eqref{eq:classicalFluxTube} is that it represents a flux-tube of
varying intrinsic thickness. The assumed form of $T(x)$, \eqref{eq:variableStringTension},
then implies that the classical flux-tube has a constant thickness
in the region $l<x<L-l$ , while the intrinsic thickness vanishes
at the end-points as shown in Fig. \ref{fig:Flux-tube}. Also, note
that the part where $T(x)$ is changing is the part where $Y_{c}(x)$
is changing, suggesting that it is the $Y$ coordinate of the string
that is related to the intrinsic thickness of the flux-tube in concordance
with the argument of \citep{Polchinski:2001ju}.

The above heuristic arguments can be made more precise for certain
confining backgrounds \citep{Witten:1998zw}. For such backgrounds
in \citep{Danielsson:1998wt} the intrinsic thickness of a very long
flux-tube was explicitly calculated. The idea being that the fundamental
string is a source of dilation field, and it is the value of dilaton
field at the boundary that couples to the square of the field-strength
tensor. Thus, by calculating the value of the dilaton field at the
boundary, which is sourced by the fundamental string, and then using
the gauge/gravity dictionary allows one to calculate $<F^{2}>$, and
obtain the transverse profile of the flux-tube. It will be interesting
to extend their calculation to a more general confining background
and for a finite string ending in a quark and an antiquark.

\section*{Acknowledgements}

I would like to thank Gunnar Bali, N. D. Hari Dass, and Rajesh Gopakumar
for very useful discussions. I am particularly grateful to N. D. Hari
Dass for his detailed comments on a preliminary version of this paper.
I would also like to thank Andreas Athenodorou, Barak Bringoltz, and
Mike Teper for a discussion on their results. This work was started
while I was visiting the Institute of Theoretical Physics, at the
University of Regensburg. I am grateful to the members of the Institute
for their hospitality, particularly to Gunnar Bali for his invitation
to visit the University of Regensburg, and DAAD for their financial
support. I have equally benefited from my visit to HRI, Allahabad,
and would like to thank Satchi Naik for his kind invitation.

\providecommand{\href}[2]{#2}\begingroup\raggedright\endgroup

\end{document}